\documentclass{elsart}

\usepackage{graphicx,epsfig}

\makeatletter

\def\elsartstyle{%

        \def\normalsize{\@setfontsize\normalsize\@xiipt{14.5}}

        \def\small{\@setfontsize\small\@xipt{13.6}}

        \let\footnotesize=\small

        \def\large{\@setfontsize\large\@xivpt{18}}

        \def\Large{\@setfontsize\Large\@xviipt{22}}

        \skip\@mpfootins = 18\p@ \@plus 2\p@

        \normalsize

}

\makeatother

\def\url#1{{\ttfamily\def\/{/\discretionary{}{}{}}#1}}

\begin{document}

\begin{frontmatter}

\title{From nuclei to atoms and molecules: the chemical history of the 
early Universe}
\subtitle{\small dedicated to Pr. Dennis William Sciama}

\author{Denis Puy$^{\dagger,\ddagger}$, Monique Signore$^\star$}
\\
\address{
$^\dagger$Institute of Theoretical Physik, University of Zurich (Switzerland)
\\
$^\ddagger$Paul Scherrer Institut, Villigen (Switzerland)
\\
$^\star$Laboratoire de Radioastronomie, Observatoire de Paris (France)}
\thanks[email]{E-mail: puy@physik.unizh.ch, monique.signore@obspm.fr}
\begin{abstract}
The {\it dark age} of the Universe is generally pointed out as the 
period between the recombination epoch ($z\sim 1000$) and the horizon of 
current observations ($z \sim 5-6$). The arrow of time in the cosmic history 
describes the progression from simplicity to complexity, because the present 
Universe is clumpy and complicated unlike the homogeneous early Universe. 
Thus it is crucial to know the nature of the constituents, in order to 
understand the conditions of the formation of the first bound objects. 
\\
In this paper we analyse the chemical history of this 
{\it dark age} through the creation of the primordial nuclei to the 
formation of the first atoms and molecules. Then 
we will describe the consequences of the molecular formation on the 
birth of the proto-objects. In this context we will mention the 
important works of Dennis 
W. Sciama who influenced a large number of theorists -cosmologists and 
astronomers- on this new field of research dedicated to primordial molecules. 
\end{abstract}
\end{frontmatter}
\section{Introduction}
At early times the Universe was filled up with an extremely dense and hot gas. 
Various important physical processes occurred as a consequence of the 
expansion. For example, 
Universe cooled below the binding energies of light elements 
leading to the formation of the nuclei which recombined with the electrons 
in order to form the first atoms. After recombination, although the density 
decrease acts against molecular formation, it turns out that the temperature 
is small enough for this formation occurs. 
\\
The existence of a significant abundance of molecules can be crucial on the 
dynamical evolution of collapsing objects. Because the cloud temperature 
increases with contraction, a cooling mechanism can be important for the 
structure formation, by lowering pressure opposing gravity, i.e. allowing 
continued collapse of Jeans unstable protoclouds. This is particularly true 
for the first generation of objects. 
\\
Interactions between primordial molecules and the cosmic microwave background 
radiation (CMBR) could be important if the molecular abundances are sufficient 
(Dubrovich 1977, Melchiorri \& Melchiorri 1994, Signore et al. 1997). 
In particular, a resonant molecular scattering between 
CMBR photons and molecules could, on one hand, smear out primary CMBR 
anisotropies and, on the other hand, produce secondary anisotropies.
\\
In this paper we will describe the chemical history of the Universe from the 
formation of the first atoms until the formation of the first objects. 
Thus in Sect.2 we will recall the nucleosynthesis 
then the important influence of the neutrinos physics in Sect. 3. The 
period of recombination will be described in Sect.4 which will lead to the 
description of the chemical chemistry or formation of primordial molecules 
in Sect.5. We will discuss in Sect.6 the influence of the molecules on 
the formation of the first objects, and on the CMBR in Sect.7. The 
possible reionization of the Universe could be some importance consequences 
on the chemical history of the Universe, in Sect.8 we will describe it, and 
we will give some conclusions and prospects in Sect.9. 
\section{Primordial nucleosynthesis}
The Standard Big Bang Nucleosynthesis model (SBBN) has been analysed first by 
Peebles (1966), Wagoner, Fowler \& Hoyle (1967), Wagoner (1969, 1973). 
Nevertheless the {\it Chicago Group} gave strong impulse to 
this field, see Yang et al. (1979, 1984), Copi et al (1995), 
Schramm \& Turner (1998), Schramm (1998) and Olive et al. (1981, 2000), 
and the {\it Oxford Pole} led by Sarkar (1996, 1999). 
Here we only summarize the main features of the SBBN model that we have 
already presented elsewhere (Signore \& Puy 1999).
\\
The standard model of the very early Universe is quite simple and 
determined by the three important milestones of the early Universe: 
the expansion governed by General Relativity, the particle interactions 
governed by Standard Model and particle distribution governed by 
statistical physics. The SBBN depends on just one parameter which 
is the baryon-to-photon ratio $\eta$ or the dimensionless quantity of 
baryon density $\Omega_b$. 
\\
In the framework of the SBBN model at times much less than one second 
after the big bang, the Universe was a hot ($T>>10^{10}$ K) rapidly 
expanding plasma, where most of its energy is on the form 
of photons at high temperature and relativistic 
particles. The weak processes such as :
$$
n+\nu \, \longleftrightarrow \, p + e^- \ , \ \
n + e^+ \, \longleftrightarrow \, p + \nu \ , \ {\rm and} \
n  \, \longleftrightarrow \, p + e^- + \overline{\nu}
$$
maintained the ratio of neutrons to protons at its thermal value, i.e. 
$n/p \sim 1$ (where $n$ define the neutrons, $p$ the protons, $e^-$ 
the electrons, $e^+$ the positrons and $\nu$ the neutrinos).
\\
At about one second, the temperature of the universe is around 10$^{10}$ K, 
the above weak processes became ineffective. The $n/p$ ratio froze out at 
about $1/6$, leading to the collisions between neutrons and protons which 
form deuterium:
$$
n + p \, \longrightarrow \, D + \gamma.
$$
The collisions with deuterium led to the formation of the helium 3 and 
tritium nuclei:
$$
n + D \, \longrightarrow \, ^3H \ , \ \ 
p + D \, \longrightarrow \, ^3He,
$$
and finally to $^4He$ via the reactions: 
$$
n +^3He \longrightarrow ^4He \ , \ \ 
p +^3H \longrightarrow  ^4He \ , \ \ 
D +D \longrightarrow ^4He.
$$
At about 100 sec. the nucleosynthesis epoch was over, thus 
most neutrons were in $^4He$ nuclei, and most protons remained free while 
smaller amounts of $D$, $^3He$ and $^7Li$ were synthetized. The low 
densities, the Coulomb barriers and stability gaps at masses 5 and 8 
worked against the formation of heavier elements. 
\\
The chemical 
composition of the Universe remains unchanged until the formation of the 
first stars. The yields of primordial nucleosynthesis are shown as a function 
of the baryon density in Fig.1 (from Burles et al. 1999).
\\
It is a considerable challenge to measure the actual primordial abundances 
because of uncertainties in measuring present day abundances and because of 
uncertainties in modeling the nuclear evolution since the big bang. The 
measured abundances of helium 4, deuterium, lithium 7 and helium 3 (Burles 
et al. 1999) are:
\begin{eqnarray}
Y_p &=&0.244 \pm 0.002 \nonumber \\
(D/H)_p &=& (3.4 \pm 0.3) \times 10^{-5} \nonumber \\
(^7Li/H)_p &=& (1.7 \pm 0.15)\times 10^{-10} \nonumber \\
(^3He/H)_p &=& (0.3 \pm 1) \times 10^{-5} \nonumber
\end{eqnarray}

Fig.1 summarizes the measured abundances with their uncertainties. Since 1980 
and until recently, cosmologists introduced a concordance interval for the 
baryon density where the predicted and measured abundances for all the four 
light elements are consistent; see for instance Copi et al. (1995) who 
derived 
$$
0.007 \, < \, \Omega_b h^2 \, < \, 0.024.
$$
A change occured in 1998-1999 with the accurate determination of the 
primordial deuterium abundance which led to the most accurate determination 
of the baryon density: Fig.1 shows the concordance intervals for each 
element and the baryon density predicted by the deuterium measurement 
(vertical band). From the latest determined primordial $D$ abundance (Tytler 
et al. 2000):
$$
(D/H)_p = (3.3 \pm 0.5) \times 10^{-5} \ \ (95\% \ {\rm c.l.}).
$$
Burles et al. (2000) give the following precise determination of the baryon 
density:
$$
\Omega_b h^2 = 0.0189 \pm 0.0019 \ \ (95 \% \ {\rm c.l.}).
$$
Until this year, the agreement between the SBBN predictions and observed 
abundances makes the SBBN a cornerstone of the Big Bang cosmology
\\
But the recent cosmic microwave background measurements of BOOMERANG and 
MAXIMA (de Bernardis et al. 2000, Jaffe et al. 2000) favor the 
following value:
$$
\Omega_b h^2 = 0.030 \pm 0.032^{+0.009}_{-0.008} \ \ (95 \% \ {\rm c.l.})
$$
which, a priori, cannot be accomodated within the SBBN model (Burles et al. 
2000). Future satellite experiments (MAP \& PLANCK surveyor) will give us 
a better cosmic microwave background determination of $\Omega_b h^2$.
\\
Meanwhile, the SBBN model seems to be in trouble... and cosmologists search 
for non-standard cosmology or non standard big bang nucleosynthesis. 

\begin{figure}[h]
\begin{center}
\epsfig{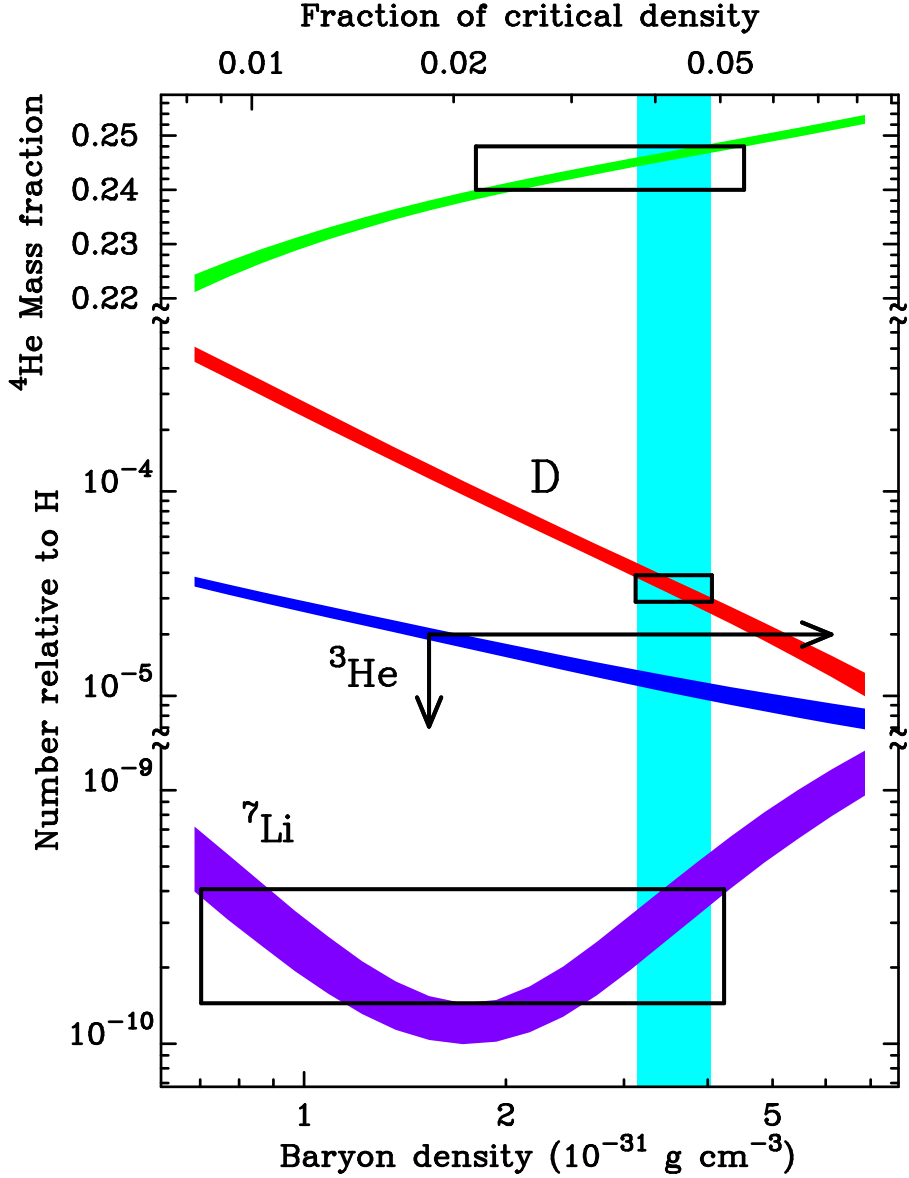}
\caption{: Relative abundances, by number: A/H, except for $^4He$ 
in mass fraction: $Y_p$. Concordance intervals for each element ($2\sigma$ 
uncertainty) and baryon density predicted (vertical line) by $D$ measurements 
are also shown (from Burles et al. 1999, {\texttt{astro-ph/9903300}}).}
\label{fig:fig1}
\end{center}
\end{figure}

\section{Neutrinos and nucleosynthesis}
In cosmology, neutrinos have a strong impact on big bang nucleosynthesis 
and therefore important consequences on the chemistry of the universe. 
Recently an extensive litterature discussed the possible 
generation of \underline{sterile neutrinos} in the early universe as a 
result of their \underline{oscillations} with \underline{active neutrinos}.
\\
Dennis W. Sciama was not only concerned with active 
neutrinos and decaying neutrinos since a long time, but also with 
sterile neutrinos. In an European conference (Sciama 1999), he gave a review 
and pointed out the importance of this field{\footnote{he argued: 
{\it students of the CMBR have now to learn something new, thanks to the 
famous announcement by the SuperKamiokande team as Neutrino 98, confirming 
the existence of an atmospheric neutrino anomaly...} (see Sciama 99)} and 
described the more recent developments on the sterile neutrinos and its 
relevance for both SBBN and the CMBR. 
\\
Thus, we briefly recall how SBBN can constrain the properties of active 
neutrinos and in particular how helium abundance can be used to count 
neutrinos.
\subsection{Limits to the number of neutrinos}
In the standard model, neutrinos possess only weak interactions, and are 
coupled to the intermediate bosons $Z^o$. Measurements of the bosonic 
decay width allow to conclude that the total number of different neutrino 
species is
$$
N_\nu \, = \, 3.07 \pm 0.12
$$
while the combined fit to LEP data gives the more accurate results:
$$
N_\nu \, = \, 2.994 \pm 0.01.
$$
In the framework of the SBBN model the precise epoch of nucleosynthesis 
was controlled by the number of particle degrees of freedom: more neutrino 
species imply a higher energy density and a faster expansion, i.e. more 
neutrons and more $^4He$ production. Therefore the $^4He$ abundance can be 
used to constrain the number of neutrinos species. For instance, Steigman 
et al. (1977) gave the SBBN limit of 7 neutrino species 
while the laboratory limit was about 5000 ! At this time, the SBBN limit was 
a very good constraint. 
\\
The helium curve in Fig.1 was computed assuming $N_\nu=3$. Burles et al. 
(2000) get the upper limit $N\nu < 3.2$ while Lisi et al. (1999) give 
$N_\nu < 4$ depending on what observational constraints one uses. 
\\
More generally, the limits on $N_\nu$ can be translated into limits on other 
types of particles (numbers or masses) that would affect the expansion 
rate just before the nucleosynthesis.
\subsection{Sterile neutrinos}
Sterile neutrinos are neutrinos which lack the standard electroweak 
interactions possessed by the active neutrinos: $\nu_e$, $\nu_\mu$ and 
$\nu_\tau$. In particular, they do not contribute to the decay width of $Z^o$; 
they would gravitate; therefore, if they existed in the early universe, their 
gravitational effects would influence the rate of expansion and therefore 
also the SBBN.
\\
Different studies have been recently done and showed 
the role played by neutrino 
oscillations in SBBN; in particular, there are some 
controversies about the 
possible generation of lepton asymmetry by oscillations between active and 
sterile neutrinos and its consequences on the observed baryon asymmetry of 
the universe, on primordial abundances (especially $^4He$ and also $D$), on 
the CMBR (Hannestad \& Raffelt 1999) and on the decaying neutrinos 
(see later in section 8). 
Look, for instance: Bell et al. (1998), Sciama (1998, 1999), Mohapatra \& 
Sciama (1998), Dolgov (2000) and references therein.
\subsection{BOOMERANG, MAXIMA and the number of neutrinos}
As already said in Sect.2, the pre-BOOMERANG and MAXIMA consensus for the 
central value for $\Omega_b h^2$ deduced from SBBN is 
$$
\Omega_b h^2 = 0.0189
$$
while the central value implied by BOOMERANG and MAXIMA measurements 
(Jaffe et al. 2000) is:
$$
\Omega_b h^2 = 0.03.
$$
The difference between the SBBN and the cosmic microwave background central 
values for $\Omega_b h^2$ has already triggered discussions, explanations 
in a lot of papers. Among them: Hu et al. (2000), Tegmark \& Zaldarriaga 
(2000), White et al. (2000), Kinney et al. (2000), Kurki-Suonio \& 
Sihvola (2000), Kurki-Suonio (2000), Hannestad et al. (2000) and 
Melchiorri \& Griffiths (2000).
\\
Let us only mention that an alternative solution can be lepton asymmetry; 
for instance, Esposito et al. (2000) find the best fit to the 
BOOMERANG and MAXIMA data with $N_\nu =9$ or suggest the best compromise 
$N_\nu =13$ for cosmic microwave background data and light elements abundances;
 Lesgourgues \& Peloso (2000) find a {\it reasonable fit} with $\Omega_b 
h^2 = 0.028$ and an effective number of neutrinos $N_{\nu}=6$. 
Active-sterile neutrino oscillations can also explain the current 
observations of cosmic microwave background and light elements abundances 
(di Bari \& Foot 2000). BOOMERANG and MAXIMA data and the number of 
neutrinos are a very active field of study.
\section{Recombination epoch}
In the hot Big Bang picture, after the nucleosynthesis period 
the natural question concerns the possibility to form atoms by recombination 
of primordial nucleus with free electrons, and the second question concerns 
the degree to which the recombination is inhibited by the 
presence of recombination radiation. 
\\
The reason for which these questions 
are central were mentionned by Peebles (1968) and Peebles \& Dicke (1968) 
based, at this epoch, on the {\it primeval-fireball} picture of Gamow (1948) 
for the evolution of the Universe. 
\\
The cosmological  recombination process was not 
instantaneous because the electrons, captured 
into different atomic energy levels, could not cascade instantaneously down 
to the ground state. Atoms reached the ground state either through the 
cosmological redshifting of the Lyman $\alpha$ line photons or by the $2s-1s$ two photons process. Nevertheless the Universe expanded and cooled faster than 
recombination could completed, and small fraction of free electrons and 
protons remained. 
\subsection{Recombination of Hydrogen}
The principles of calculations of the primordial recombination have been 
mentionned initially by Shklovskii (1967), Novikov \& Zel'dovich (1967) and 
Takeda \& Sato (1968). Peebles (1968) was the first to 
present a theory in which the very complicated recombination process is 
reduced to simpler terms\footnote{{\it by means of a number of approximations valid in 
the conditions of interest, so that the physical situation can be described by a few variables, and the recombination equation can be integrated  by a few 
variables, and the recombination equation can be integrated by hand, without 
having to abandon any of the essential elements of the problem} as written 
by Peebles himself in 1968.}. The methodology of the calculations consist to 
take into account a three level atom with a ground state, first excited state and continuum, represented by a recombination coefficient. A single ordinary 
differential equation is derived to describe the ionisation fraction 
(see Peebles 1968). The assumptions are:
\begin{itemize}
\item Hydrogen excited states are in equilibrium with the radiation
\item Stimulated de-excitation is negligible
\item Collisional processes are negligible
\end{itemize}
The approximations are based on the rate of transfer of energy per unit 
volume between radiation and free electrons 
developped by Kompaneets (1957) then Weymann (1965, 1966) and coupled with 
the expansion of the Universe. 
\\
Peebles (1968) computed the rate of recombination of the plasma from the rate 
coefficient for recombination to excited states of hydrogen tabulated by 
Boardman (1964), and showed that the residual ionization of the hydrogen 
(or fractional ionization $x_e$) is below 5.3 $\times$ 10$^{-5}$ for a flat 
       cosmological model at redshift below 1500. 
Jones \& Wyse (1985) developped two important analytic approximations: 
one, valid at the redshift $z>1500$ is a modification of the standard Saha 
equation, and the other, valid for $800<z<1500$. They found that the 
fractional ionization is roughly 6.75 $\times$ $10^{-4}$ at the redshift $z=555$ 
much higher than the Peebles values. More recently Seager, 
Sasselov \& Scott (1999, 2000) have presented a refined treatment of the 
recombination through a complete code which performs appproximate 
calculations history\footnote{see 
{\texttt{http://www.astro.ubc.ca/people /scott/recfast.html.}} }.

\subsection{Recombination of Helium}
We have seen that $^4He$ is the second most abundant element in the Universe 
after hydrogen. Zel'dovich, Kurt \& Sunyaev (1968) and Matsuda, Sato \& 
Takeda (1969) were the first to point out 
that the cosmological recombination of helium differs from that of 
hydrogen. The conditions for the helium recombination are such that in both 
cases it occurs in accordance with the Saha equation 
(Bernshtein \& Dubrovich 1977, Dubrovich \& Stolyarov 1997). 
Recombination history of early Universe consist of three stages but 
helium recombination occurs in two steps:

\begin{itemize}
\item at $z\sim 6000$ (or when the temperature became less than 16 000 K) 
the first electron recombines and singly ionized $He$II 
forms from doubly ionized $He$III.
\item Then at $z\sim 2700$ (or temperature close to 7300 K), $He$II 
recombines into a neutral state.
\item from $z \sim 1500$ recombination of hydrogen took place.
\end{itemize} 
More recently Seager, Sasselov \& Scott (2000) have developped an improved 
recombination calculation of $H$, $He$ (with $He$II and $He$III) that 
involves a line-by-line treatment of each atomic level. They found that $He$I 
recombination is much slower than previously thought, and it is delayed until 
just before $H$ recombines. 
\subsection{Recombination of Lithium}
The ionization potential $I_{Li}=5.392$ eV of $Li$I is shorter than that of 
hydrogen ($I_H = 13.6$ eV). Dalgarno \& Lepp (1987) suggested the existence of 
$Li^+$ ions after the period of the  hydrogen recombination, and showed that 
the radiative recombination is again effective at the redshift $z\sim 450$. 
This last point is particularly important for the lithium chemistry as we 
will see.
\section{Primordial chemistry}
From the pioneer works of Zwicky (1959) on the molecular hydrogen content 
of the Universe, presented at the San Francisco meeting of the 
Astronomical Society of the Pacific in June 1959, 
the idea of molecular formation at the immediate 
post-recombination took form. 
\\
The studies on the primordial chemistry (or post-recombination chemistry) 
have been the source of a tremendous increase of the litterature. 
The first complete description of the hydrogen chemistry was developped by 
different japanese groups: Hirasawa, Aizu \& Taketani (1969), Takeda, Sato 
\& Matsuda (1969) and Matsuda, Sato \& Takeda (1969) in the context of 
pre-galactic gas clouds or formation of galaxies. The importance of the  
deuterium chemistry through the formation of the $HD$ molecule was suggested 
by Lepp \& Shull (1983) as lithium chemistry through the formation of the 
molecule $LiH$, completed by the works of Dalgarno \& Lepp (1987) on 
the helium chemistry and the lithium chemistry. Latter a complete 
comprehensive chemistry of the lithium, at the post-recombination epoch, 
was presented 
by Stancil, Lepp \& Shull (1996). 
\\
The first chemical network including 
the primordial molecules (such as $H_2$, $HD$ and $LiH$) and ions was carried 
out by Lepp \& Shull (1984), Latter \& Black (1991), Puy et al. (1993) and 
more recently by Galli \& Palla (1998), and Stancil, Lepp \& Dalgarno (1998). 
The development of the primordial chemistry studies owe to 
Dalgarno an important contribution concerning the calculations of the 
reaction rates (see Lepp \& Shull 1983, Dalgarno \& Lepp 1987, Puy et al. 1993 
and Galli \& Palla 1998 and references therein).
\subsection{Hydrogen chemistry}
After the important works of the recombination period, it was plausible to 
imagine that molecular hydrogen plays a role in the early evolution of large 
clouds. Nevertheless $H_2$ was not considered as a component of the 
pre-galactic medium. 
\\
In a review article dedicated to the hydrogen molecules 
in astronomy Field, Somerville \& Dressler (1966), showed that the times 
scales for appreciable amounts to form in three-body reactions or in 
radiative association were always greater than the Hubble age. In the 
cosmological context there are no grains which can catalyze the reaction of 
formation, thus any $H_2$ formed in the uniform background is dissociated 
by the radiation, until the density is too low to produce it. 
Saslaw\footnote{In 1967, William C. Saslaw was a Ph.D. student of 
Dennis W. Sciama at the Department of Applied Mathematics and 
Theoretical Physics of the University of Cambridge (see the {\it family tree} in Ellis et al. 1993)} \& Zipoy (1969) then Shchekinov \& Ent\'el (1983) 
showed the importance of the charge 
transfer reactions: 
\begin{equation}
H_2^+ + H \, \rightarrow \, H_2 + H^+
\end{equation}
initiated by the radiation association 
\begin{equation}
H + H^+ \, \rightarrow \, H_2^+ + h\nu. 
\end{equation}
They pointed out that $H_2^+$ ion is converted to $H_2$ as soon as it is 
formed and the $H_2^+$ concentration never becomes large. 
\\
Peebles \& Dicke (1968), in a scenario concerning the origin of the 
globular star clusters, showed that these clusters  may have originated as 
gravitationnally bound gas clouds before the galaxies formed, and suggested 
that some molecular hydrogen can form, mainly by way of negative 
hydrogen by the reactions (see also Shchekinov \& Ent\'el 1983)
\begin{equation} 
H + e^- \, \rightarrow \, H^- + h\nu
\end{equation}
followed by the reaction
\begin{equation}
H^- + H \, \rightarrow \, H_2 + e^-
\end{equation}
with the reaction rates calculated by Mc Dowell (1961) from remarks 
on the quantum theory of the negative hydrogen ion emphasized by 
Chandrasekhar (1944), in which the electrons act only as catalysts. 
Thus two possible ways of formation were pointed out. 
\\
Takeda, Sato \& Matsuda (1969) were the firsts to study the evolution of 
molecular hydrogen abundance in the cosmological medium 
(i.e. post-recombination Universe) in contrast with the 
works by Saslaw \& Zipoy (1967) and Peebles and Dicke (1968) in which they 
calculated the products of $H_2$ in dense clouds. They considered the two ways 
of $H_2$ formation and electron and proton as a kind of catalyzer. 
\\
The photodetachment of $H^-$ and the photodissociation of $H_2^+$ by the 
background radiation field restrict the abundance of molecular hydrogen formed 
at early stages, although the photodestruction of molecular hydrogen is 
negligible. The destruction is due to collisional dissociation 
(Lepp \& Shull 1983). 
\subsection{Deuterium chemistry}
Deuterium chemistry in a early Universe could play an important role in the 
sense that it could give some explanations on the controversy observations 
of fractional abundance in high redshift Ly$\alpha$ (Songaila et al. 1994, 
Burles \& Tytler 1996).  
Although similar processes contribute to the formation of $HD$:
\begin{equation}
D^+ + H \, \rightarrow \, HD^+ + h\nu
\end{equation}
\begin{equation}
HD^+ + H \, \rightarrow \, H^+ + H_2,
\end{equation}
its formation also proceeds through:
\begin{equation}
H^+ + D \, \rightarrow \, D^+ + H
\end{equation}
\begin{equation}
D^+ + H_2 \, \rightarrow \, H^+ + HD.
\end{equation}
Thus the formation of $HD$ is carried out in two steps. Nevertheless when the abundance 
of $H_2$ is sufficient, the second way of formation is dominant (Eq. 8). 
Stancil, Lepp \& 
Dalgarno (1998) presented a complete review of the deuterium chemistry of the 
early Universe. 
\subsection{Helium chemistry}
The presence of helium in the early Universe gives rise to a rich assembly of 
molecular processes as pointed out by Hirasawa (1969). The main molecular species containing helium is $HeH^+$. Processes that lead to the formation and 
destruction of the molecular ion $HeH^+$ in astrophysical environments have 
been the subject of a number of investigations. Dabrowski \& Herzberg (1978) 
were the one of the firsts to introduce the possibility that this 
molecular ion exists in astrophysical plasmas; it is from these initial 
works that Roberge \& Dalgarno (1982) presented the formation 
and destruction mechanisms.
 More recently Zygelman, Stancil \& Dalgarno (1998) calculated the 
enhancement of the rate coefficient for the radiative association to form 
$HeH^+$:
\begin{equation}
He + H^+ \, \rightarrow \, HeH^+ + \nu.
\end{equation}
The $H_2^+$ ions and $H_2$ produce $HeH^+$ by the fast reactions
\begin{equation}
H_2^+ + He \, \rightarrow \, HeH^+ + H,
\end{equation}
\begin{equation}
H_2 + He^+ \, \rightarrow \, HeH^+ +H.
\end{equation}
Nevertheless for all of these processes, once photodissociation ceases to be 
rapid, the $H_2^+$ ions react preferentially with neutral atomic hydrogen to 
form molecular hydrogen as we have seen.
\subsection{Lithium chemistry}
The lithium chemistry is initiated by the recombination of lithium, 
which occured near 
the redshift $z \sim 450$. The formation of the molecular ion $LiH^+$ formed by radiative 
association processes:
\begin{equation}
Li^+ + H \, \rightarrow \, LiH^+ + h\nu,
\end{equation}
\begin{equation}
H^+ + Li \, \rightarrow \, LiH^+ + h\nu
\end{equation}
opens the way of the formation of the $LiH$ molecules through the 
exchange reactions
\begin{equation}
LiH^+ + H \, \rightarrow \, LiH + H^+,
\end{equation}
which are more rapid than the formation by radiative association of 
$H$ and $Li$ atoms. 
\\
The complete description of the lithium chemistry was emphasized by Stancil, 
Lepp \& Dalgarno (1996) where new quantal rate coefficients were included. 
\subsection{Molecular abundances}
In order to estimate the molecular abundances which depend on 
the reaction rates\footnote{
the complete chemical network is described in Puy et al. (1993), 
Galli \& Palla (1998) and Stancil et al. (1998).}, it is necessary to know 
the thermal and dynamical evolution of the medium. 
In the Einstein-de Sitter 
Universe the evolution of matter temperature $T_m$ is 
described by the equation:
\begin{equation}
\frac{d T_m}{dt} \, = \, \alpha x_e T_r (T_r - T_m) - \frac{2 T_m}{a} \, 
\frac{da}{dt}
\end{equation} 
where $\alpha$ is a constant, $x_e$ the fractional ionization of the 
hydrogen, $T_r$ the temperature of the cosmological radiation temperature 
and $a(t)$ the expansion parameter. The first term describes the heat transfer 
from radiation to the electrons when the second term characterizes the 
cooling due to the expansion of the Universe. The result of the integration 
of this important equation led to the idea that there is thermal decoupling 
between matter and radiation.
\\
Nevertheless the formation of primordial molecules such as $H_2$, $HD$, and 
$LiH$ can involve a thermal response due to the excitation 
of the rotational levels  of these molecules. The population of the rotational 
levels is mainly due to collisional excitation and de-excitation with $H$, 
$H_2$ and $He$ on one hand and to radiative processes on the other hand: 
absorption from the 
Cosmic Microwave Background Radiation (CMBR) and spontaneous or induced 
emission. 
\\
The importance of radiative cooling rate by molecular hydrogen were pointed 
out by 
Takayanagi \& Nishimura (1960), which developped theoretical calculations of 
the cooling process due to the rotational excitation of $H_2$ molecule. They 
considered  collisions with hydrogen atom in the context of interstellar 
clouds. Saslaw \& Zipoy (1967) used the radiative function of Takayanagi 
\& Nishimura (1960) in the context of pre-galactic gas clouds. 
The molecules $HD$ and $LiH$ play an important thermal 
role, since their allowed dipole rotational transitions, see 
Lepp \& Shull (1984). 
Puy et al. (1993) calculated precisely the population of the levels and 
provide a molecular thermal function $\Psi_{mol}$ due to $H_2$, $HD$ and 
$LiH$. A complete description of the cooling of astrophysical media 
is found for $H_2$ in Le Bourlot et al. (1999), and for $HD$ in Flower et 
al. (2000). Flower (2000) presented recently a study on the role of $HD$ 
in the thermal balance of the primordial gas and conclude that $HD$ is 
about as important as $H_2$ in the thermal balance of the primordial gas. 
Notice that the estimation of the ortho-para $H_2$ in the 
primordial gives a better estimation of the $H_2$ cooling as Flower \& Pineau 
des For\^ets (2000) emphasized very recently. 
Furthermore some chemical reactions 
produce a chemical thermal function $\Theta_{chem}$ through the enthalpy of 
reaction. 
\\
In the cosmological context the thermal evolution equation becomes
\begin{equation}
\frac{d T_m}{dt} \, = \, \alpha x_e T_r (T_r - T_m) - \frac{2 T_m}{a} \, 
\frac{da}{dt} 
+ \frac{2 \Psi_{mol}}{3 n k} + \frac{2 \Theta_{chem}}{3 nk}
\end{equation}
Puy et al. (1993) took into account the transfer process between radiation and 
matter via the Thomson scattering, coupled with the molecular source term and 
the enthalpy of the reactions in order to calculate the abundances of the 
species\footnote{The numerical integration of the coupled chemical equation 
is an initial value problem for stiff differential equations 
(see Puy et al. 1993).}. 
\\
After a transient growth all abundances of the molecules become almost 
constant. The final abundances of $H_2$, $HD$ and $LiH$ {\it freeze out} 
due to the expansion. Moreover the final values of molecular abundances 
depend on the choice of the cosmological parameters. Galli \& Palla 
(1998) summarized and discussed the chemistry of early Universe for 
different cosmological models.
\\
In the standard model defined by the following cosmological parameters:
\\
\\
$H_o=67$ for the Hubble constant and $\eta_{10}=4.5$ for the baryon-to-photon 
ratio in an Einstein-de Sitter Universe. In this context we have at the 
end of the recombination, the following initial abundance:
\\
\\
$D/H \sim 4.3 \times 10^{-5}$, and $Li/ H \sim 2.2 \times 10^{-10}$.
\\
\\
In this context we obtained the following final abundance at $z=5$:
\\
\\
$H_2/H \sim 1.1 \times 10^{-6}$, 
$HD/H \sim 1.2 \times 10^{-9}$ and 
$LiH/H \sim 7.2 \times 10^{-20}$
\\
\\
and for the main molecular ions with $x_e \sim 3 \times 10^{-4}$:
\\
\\
$H_2^+/H \sim 1.3 \times 10^{-12}$, 
$HD^+/H \sim 2.1 \times 10^{-18}$, 
$H_2D^+/H \sim 5.1 \times 10^{-14}$, 
\\
$HeH^+/H \sim 6.2 \times 10^{-13}$ and 
$LiH^+/H \sim 9.4 \times 10^{-18}$.
\section{Primordial objects} 
The developement of structure in the universe was well advanced at high 
redshifts. For example, quasars have been detected nearly to $z\sim 5$, and 
the most distant galaxies to even greater distances. 
Still unknown, however, is the epoch during which the 
first generation of objects was formed. Hoyle (1953) were one of the firsts 
to suggest that the importance of the 
fragmentation process of extragalactic hydrogen cloud gas clouds, 
with particular reference to the formation of galaxies out of hot hydrogen 
clouds. From these pioneer works Hunter (1962) developped an instability 
theory of a collapsing gas cloud.
\\
Gravitational instability and thermal instability were supposed to be the main 
processes to form condensations in a dilute gas. In the framework of the 
gravitational instability theory, each structure started as a tiny local 
overdensity; nevertheless very little is known about the protoclouds. 
Fluctuations that survive decoupling are subject to gravitational 
instabilities if they are on sufficiently large scale. The general 
treatment of the gravitational instability in an expanding Universe, developped 
by Lifshitz (1946) and Lifshitz \& Khalatnikov (1963), showed that a 
condensation due to gravitational instability cannot grow so fast during the time 
scale of the Universe. Parker (1953) suggested that the problem of condensation 
can be understood as a consequence of instability on the thermal equilibrium of a 
diffuse medium. 
\\
Since the works of Jeans (1902) the importance of thermal 
instability as another possible mechanism for galaxy formation were made by Hoyle 
(1958) and above all Field (1965). Kato, Nariai \& Tomita (1967) examined the 
thermal instability of a dilute gas in expanding Universe. Sofue (1969) found that 
formation of galaxies could be 
initiated by thermal instability due to radiative 
cooling, and developped by Saito (1969) in the context of a uniformly 
rotating homogeneous medium. The formalism were definitively established 
by Kondo (1970) in an expanding medium. 
\\
Doroshkevish, Zel'dovich \& Novikov (1967) showed that if the cloud 
contracts adiabatically, hydrogen atoms in it are soon collisionally ionized again 
and the cloud is finally dispersed by radiation pressure. Therefore, some kind of 
cryogen is necessary to form a bound system as mentionned Takeda, Sato \& Matsuda 
(1968). The problem was to find a good cryogen and so primordial molecules offered 
an interesting solution. Matsuda, Sato \& Takeda (1969) realized that hydrogen 
molecule could be a possible cooling of pre-galactic gas clouds from the 
works on the $H_2$ cooling in the interstellar clouds of Takayanagi \& 
Nishimura (1960). Yoneyama (1972) pointed out that after hydrogen molecules 
are formed, the cooling of the gas by these molecules plays an 
important role on the further condensation and fragmentation of the cloud.
\\
Since these first works, the hypothesis that galaxies formed from density 
perturbations which collapse within the expanding background of a Friedmann 
universe appears worthy of continued investigation. Nevertheless all of these 
first models considered the collapse of non-rotating spherical clouds. Thus 
Hutchins (1976) studied the thermal effects of $H_2$ molecules in rotating and 
collapsing spheroidal gas clouds, and suggested an early period of formation 
of objects in the mass range of ordinary stars; which led to the model 
of Silk (1977) concerning the fragmentation of cosmic gas clouds 
consisting of gas of a predominantly primordial composition. The contraction 
of a such cloud is initially adiabatic. 
\\
Early studies focused on the chemical evolution and cooling of primordial 
clouds by solving a chemical reaction network within highly idealized 
collapse models (Palla et al. 1983, Mac Low \& Shull 1986, Puy \& Signore 
1996, Tegmark et al. 1997). Some hydrodynamic aspects were studied in 
spherical symmetry and multidimensional studies by Anninos \& Norman (1996) 
and more recently by Abel et al. (2000). 
\\
A small fractional abundance of $H_2$ 
molecules is formed via the $H^-$ way. The small residual ionization 
remaining after decoupling is sufficient to produce enough $H_2$ to radiate 
away the compressional energy of collapse. Further collapse proceeds almost 
isothermally, and considerable fragmentation occurs. This process is 
terminated only when at sufficiently high densities the $H_2$ molecules are 
destroyed by the reaction:
\begin{equation}
H_2 + H + H \, \rightarrow \, 3H.
\end{equation}
Palla, Salpeter \& Stahler (1983) examined precisely the three body reactions
\begin{equation}
H+ H + H \, \rightarrow \, H_2 + H
\end{equation}
\begin{equation}
H + H + H_2 \, \rightarrow \, H_2 + H_2,
\end{equation}
and investigated the thermal and chemical evolution of a collapsing 
spherical cloud composed of pure hydrogen gas, and various regimes 
of oscillation in the collapse can occurs as Lahav (1986) showed.
\\
Lepp \& Shull (1984) emphasized that the dipole rotational transitions of $HD$ and $LiH$ are particularly important at high density and low temperature. 
In this context Puy \& Signore (1997) examined the evolution of primordial molecules 
in a context of gravitational collapse and showed how the abundances 
can be modified. The importance of $HD$ molecules was pointed out by Puy \& 
Signore (1998b) where they analysed the ratio between the molecular cooling 
due to $HD$ and that due to $H_2$, and found that the main cooling agent 
around 200 K is $HD$. This results was confirmed by Okumurai (2000), 
Uehara \& Inutsuka (2000) and Flower et al. (2000).
\\
The cooling of $LiH$ is less important. Lepp \& Shull (1984) and 
Puy \& Signore (1996) considered the $LiH$ molecules formed through only 
the radiative association. Stancil, Lepp \& Dalgarno (1996), Dalgarno, 
Kirby \& Stancil (1996) and Gianturco \& Gori-Giorgi (1996) proposed a new 
lithium chemistry with new 
ways of $LiH$ formation. In this context Bougleux \& Galli (1997) then 
Puy \& Signore (1998a) developped a complete lithium chemistry in a 
collapsing cloud. Nevertheless at the temperature higher than 200 K, $H_2$ 
and  $HD$ cooling are thermically dominant as Abel et al. (1997) 
pointed out in three-dimensional numerical simulations, 
which lead to a scenario of the formation of first stars in the Universe 
due to the fragmentation of primordial gas (see also Bromm, Coppi \& Larson 
1999).
\\
Thus primordial molecules play an important role in the pregalatic gas 
particularly on the thermochemical stability as mentionned by 
Rozenzweig et al. 
(1994) and on the problem of the minimum mass which virialized in order to 
form the first cosmological objects (Tegmark et al. 1997).
\section{Anisotropies of CMBR and primordial molecules}
The study of the CMB anisotropies (CMBA) -primary and secondary ones- are 
important tools to study the origin and the evolution of perturbations. 
Different physical processes are responsible for the production of the 
CMBA: each one associated with a typical angular scale. 
\\
Among the different kinds of secondary CMBA, the Rees-Sciama effect (Rees \& 
Sciama 1968) 
can produce a detectable signal only if very large masses are involved 
($M>10^{16}$ M$\odot$) and can give significant contributions only at large 
angular scales (degree scale) and in $\Omega \neq 1$ scenarii. The 
Sunyaev-Zel'dovich effect 
(Sunyaev \& Zel'dovich 1972) is important at cluster scales 
(subarcminute scale). At very small scales (subarcminute scale) two kinds 
of secondary CMBA can be sizeable: thermal emission by dust in galaxies and 
primordial molecular lines produced by resonant elastic scattering. Here, we 
only consider the primordial molecular lines.
\subsection{Interaction between the CMB and primordial molecules}
From an initial idea of Zel'dovich, Dubrovich (1977) showed that resonant 
elastic scattering must be considered as the most efficient process in 
coupling matter and radiation at high redshift. They noted that the cross 
section for resonant scattering between CMB and molecules is several orders 
of magnitude larger than that between CMB and electrons: even a modest 
abundance of primordial molecules would produce significant Thomson scattering.
 At this point, every velocity field (due to molecular motion or to cloud 
infall) would leave its imprint on CMB via Doppler shift. This technique 
for exploring the early universe has been revised by the group of 
Melchiorri (De Bernardis et al. 1993, Melchiorri \& Melchiorri 1994, Maoli 
et al. 1994, Signore et al. 1994). A careful analysis of various primordial 
molecules has led to the identification of $LiH$ and $LiH^+$ as the best 
candidates (Dubrovich 1977, de Bernardis 1993). Note only that this 
conclusion was relevant because all these authors assumed a value for the 
$LiH$ abundance (i.e. $LiH/H \sim 3\times 10^{-8}-3\times 10^{-9}$) which 
was adopted -at that time- by all the experts in primordial chemistry.
\\
An attempt to search for $LiH$ has even been carried out with the IRAM 30m 
telescope (de Bernardis et al. 1993). Because the resonant scattering is an 
elastic process, it can result in a primary CMBA attenuation and in a 
secondary CMBA production, exactly as in the case of an early reionisation of 
the universe. Note that this effect is strongly frequency-dependent while the 
effect of Thomson scattering in a ionized medium is not frequency dependent. 
Finally, one can show that the blurring of primary CMBA by resonant scattering lines of $LiH$ can occur if $LiH/H > 3 \times 10^{-8}$ for 
$\nu < 60$ GHz and for intermediate angular scales (see Maoli et al. 1994).
\subsection{Possible molecular signals ?}
Just as primary CMBA could be erased, secondary CMBA could be expected by 
molecular resonant scattering. In the framework of a Cold Dark Matter (CDM) 
scenario -and also in a more general case- Maoli et al. (1996) have 
calculated the intensity the line width of the expected signal during the 
three stages of the evolution of a very simple protocloud model: linear 
phase, turn-around epoch and non-linear collapse. In particular, they have 
shown that for standard observational conditions (10$^{-5}<\Delta \nu / \nu < 
10^{-4}$; $\Delta I / I_{CBR} > 10^{-4}$ the beginning of the non-linear 
collapse phase of a protocloud, just after its {\it turn-around} epoch, is the 
best one for a detection of the first two $LiH$ rotational lines which are 
observable in the frequency range $30<\nu<250$ GHz with an angular scale range 
7''$<\theta<20$''. They concluded that the IRAM 30-m telescope with its 10'' 
of angular resolution and its frequency channels, is well suited for this 
kind of observations, see also Signore et al. (1997).
\subsection{Discussion}
The effects of $LiH$ molecules on the CMBA -erasing of primary CMBA, creation 
of secondary CMBA- strongly depend on the final $LiH$ abundance which is a 
function of the lithium produced by primordial nucleosynthesis and of the 
efficiency of its conversion to $LiH$. With the quantal rate coefficient  
for the radiative association of lithium and hydrogen given by 
Stancil et al. (1996) (see Sect.6) the fractional abundance of $LiH$ 
is much smaller than that of the above CMBA studies and therefore leads to 
\underline{no erasing of the primary CMBA} and \underline{no production of 
secondary CMBA}. But let us only emphasize that, again at present times, the 
primordial lithium abundance and the percentage of lithium converted to 
$LiH$ are both quite uncertain !
\section{Reionization}
In the 1960's years there were several ways which led to the idea that 
the medium at high redshift could be ionized: The radiation from a very 
distant object could be scattered by free electrons (Field 1954, 
Gunn \& Peterson 1965); radio waves of low frequency from extragalactic 
sources would be absorbed (Field 1954, Sciama 1964a, Ericson \& Cronyn 1965),
 and the radiation from radio sources could suffer dispersion (Haddock and 
Sciama 1965). Sciama (1964a, 1964b) and thereafter Rees \& Sciama (1966) were 
one of the first to put forward that the intergalactic hydrogen could be 
partially ionized at large scale.
\\
The absence of a 
Gunn-Peterson (Gunn \& Peterson 1965) effect in the 
spectra of high redshift quasars 
implies that the intergalactic medium is highly ionized at low 
redshifts (i.e. $z\sim 5$). These observations suggested that the 
intergalactic medium was reionized in the redshift interval 
($5<z<300$)\footnote{The detection of CMBR anisotropies rules out a fully 
ionized intergalactic medium beyond $z\sim 300$, see Scott, Silk \& White 
(1995)}. Thus at epochs corresponding to $z\sim 1000$ the intergalactic 
medium is expected to recombine and remain neutral until sources of radiation 
and heat develop that are able of reionizing it. 
\\
A number of suggestions have been made for possible alternative 
sources of intergalactic ionizing photons. Sciama (1993a) gave an exhaustive 
description of the scenario of these alternatives sources.
\\
The scenario through decay of dark matter particles has been proposed 
by different authors, the first was pointed by Cowsik (1977) and extented by 
de Rujula \& Glashow (1980) from massive neutrinos decayed into lighter 
neutrinos and UV photons. The hypothesis of the photoionization of the 
intergalatic medium by radiatively decaying neutrinos or photinos has been 
previously formulated by Sciama (1982a, 1982b, 1982c), which led also to a 
scenario describing the ionization of $HI$ regions in the galaxy 
(Sciama 1990a) or the ionization of Hydrogen throughout the Universe (Sciama 
1990b, 1993b, 1994, Sethi 1997).
In this context 
Salati \& Wallet (1984) reinvestigated the recombination of neutral hydrogen 
taking into account light neutral fermions, stable or radiatively unstable. 
They found a limit on the fractional ionization: 
\\
$
4 \times 10^{-4} \, < \, x_e \, < \, 2 \times 10^{-3}
$ 
\\
instead of the previous baryon-dominated universe result: 
\\
$3 \times 10^{-5} \,  < \, x_e \, < \, 3 \times 10^{-4}$.
\\
Sciama (1988) from the works of Cabibbo, Farrar, Maiani (1981), 
Sciama \& Melott (1982) and Sciama (1982c, 1984) proposed a scenario of photino 
decay in order to explain the observation of the ionization of Lyman-$\alpha$ 
clouds at large redshifts.
\\
Bowyer et al. (1999) observed the spectrum of the night sky 
with an extreme-ultraviolet spectrometer covering the bandpass from 350 to 
1100 $\AA$ on board the spanish satellite {\it MINISAT-1} launched in 1997 
April 21. The observed is far less intense than that expected in the Sciama 
model of radiative decay of massive neutrinos, but they cannot 
exclude the earlier model of Melott (1984), based also on the theory of 
Sciama with a decay energy which is somewhat greater than 13 eV, and 
different than the 13.6 KeV of the Sciama model.
\section{Conclusions}
First, let us recall that the latest CMB results require a higher baryon 
density than allowed by SBBN theory and observations and therefore new physics 
-for example a strong lepton asymmetry-
\\
Then, we have seen that the chemistry could have consequences on the 
formation of 
the proto-objects. Nevertheless many important questions still await answers. 
In primordial chemistry a lot of reaction rates are estimated only 
and particularly those involved in 
the formation of $LiH$ molecules. Recently Dickinson \& Gadea (2000) 
developped a fully quantal description of the radiative association in 
$Li^+ + H$ collisions, and showed that the rate coefficient is at least 
five orders of magnitude larger than that for the classical radiative 
association $Li+H$. Thus the radiative cooling due to $LiH$ could be changed 
($LiH$ has a strong dipole moment).
\\
Calculation of the rate coefficients, for rotational transitions induced 
in collisions between the primordial molecules and atoms, is crucial. For 
example Roueff \& Zeippen (1999) presented calculations of rotational 
excitation of $HD$ molecules by $He$ atoms, and provide a new estimation of 
molecular cooling and heating. In this context  Le Bourlot et al. (1999) 
and Flower et al. (2000) showed the importance to use the recent quantum 
mechanical calculations of cross-sections for rotational transitions, in 
order to estimate respectively the $H_2$ cooling and $HD$ cooling. 
Thus recently the rate 
coefficients are accessed in the software library of {\it Collaborative 
Computational Project}\footnote{{\texttt{http://ccp7.dur/ac.uk/}}}, 
the {\it UMIST} data-base
\footnote{{\texttt{http://saturn/ma.umist.ac.uk:8000/}}} which 
provide a large data of reaction rates and collisional coefficients. 
Primordial chemistry is still in a nascent stage: the development of 
computing technics and specifical software for the quantum mechanics 
will open large possibilities. 
\\
Primordial molecules could play a role at the turn-around period, i.e. 
where the expansion of the density perturbation is maximum and begins to 
collapse, and a modification of the temperature at this turn-around point can 
occur (Puy 2000b, Puy \& Signore 2001). Moreover the primordial molecules 
play a crucial role on the evolution of the first objects, through the 
process of fragmentation. The first objects could be massive 
stars. Thus, as soon as stellar processes occur, proto-objects can lead 
through SN explosions to the formation of other molecules species, such as 
$CO$, $CI$ or $HCN$. They in turn are important sources of 
contamination of the medium, and thus can offer different ways of early 
pre-biotic molecular formation. Chakrabarti \& Chakrabarti (2000) showed that 
a significant amount of adenine ($H_5C_5N_5$), a DNA base, may be produced 
during molecular cloud collapse through the chain reaction of $HCN$ addition. 
The existence of primordial carbon or nitrogen could also produce primordial 
bio-molecules or pre-biotic molecules just after the cosmological 
recombination period (see Puy 2000a). 
\\
The researches and the perpectives on primordial molecules 
are very large. The ESA cornerstone mission as Far InfraRed and Submillimetre 
Telescope (FIRST), or the Atacama Large Millimetre Array project (or ALMA), 
with a collecting area of up 10 000 m$^2$, will offer interesting perspective 
of detection.

\section*{Acknowledgements}
One of the authors (D.P.) was EEC Marie Curie Post-Doc Fellowship (1994-1996) 
at the International School of Advanced Studies (SISSA-Trieste) on the 
supervision of Dennis W. Sciama. 
Dennis W. Sciama gave complete freedom to choose the topics, and 
always followed this field with constant support and encouragement. 
The multiple discussions with Dennis W. Sciama 
showed that a great physicist have not only a formidable knowledge of the 
physics, but also a real sense of the human respect.
\\
This works have been supported by the {\it Dr Tomalla Foundation} and by the 
Swiss National Science Foundation.
\vskip4mm
\begin{center}
\vskip2mm
\textsl{\large \it He was a man, take him for all in all.
\\
I shall not look upon this like again...}
\\
(Hamlet, William Shakespeare)
\end{center}
\section*{References}
{\footnotesize
Abel T., Anninos P., Zhang Y., Normal M. 1997 New Ast. 3, 181
\\
Abel T., Bryan G., Norman M. 2000 ApJ 540, 39
\\
Aninos P., Norman M. 1996 ApJ 460, 556
\\
Bell N. et al. 1998 Phys. Rev. D58, 5010
\\
Bernshtein I., Bernshtein D., Dubrovich V. 1977 Astron. Zh. 54, 727
\\
Boardman W. 1964 ApJ Suppl. 9, 185
\\
Bougleux E., Galli D. 1997 MNRAS 288, 638
\\
Bowyer S, Korpela E., Edelstein J. et al. 1999 ApJ 526, 10
\\
Bromm V., Coppi P., Larson R. 1999 ApJ 527, L5
\\
Burles S., Tytler D. 1996 ApJ 114, 1330
\\
Burles S., Nollett K., Turner M. 1999, {\texttt{astro-ph/9903300}} 
\\
Burles S., Nollett K.,Turner M. 2000, {\texttt{astro-ph/0008495}} 
\\
Cabibbo N., Farrar G., Maiani L. 1981 Phys. Lett. B 105, 155
\\
Chakrabarti S., Chakrabarti S.K. 2000 A\&A 354, L6
\\
Chandrasekhar S. 1944 ApJ 100, 176
\\
Copi C., Schramm D., Turner M. 1995 Science 267, 192
\\
Cowsik  R. 1977 Phys. Rev. Lett. 39, 784
\\
Dabrowski I., Herzberg G. 1978 Ann. NY Acad. Sci. 38, 14
\\
Dalgarno A., Lepp S. 1987 in {\it Astrochemistry}, ed. Vardya and Tarafdar, 
Dordrecht-Reidel Eds
\\
Dalgarno A., Kirby K., Stancil P. 1996 ApJ 458, 397
\\
De Bernardis P. et al. 1993 A\&A 271, 683
\\
De Bernardis P. et al. 2000 Nature 404, 955
\\
De Rujula A., Glashow S. 1980 Phys. Rev. Lett. 45, 942
\\
Di Bari P., Foot R. 2000 {\texttt{astro-ph/0008258}}
\\
Dickinson A., Gadea F. 2000 MNRAS 318, 1227
\\
Dolgov A. 2000 hep-ph/0004032
\\
Doroshkevich A., Zel'dovich Ya., Novikov I. 1967 Soviet Aj. 11, 233
\\
Dubrovich V. 1977 Sov. Astr. Lett. 3, 128
\\
Dubrovich V., Stolyarov V. 1997 Astron. Lett. 23, 565 
\\
Ellis G., Lanza A., Miller J., in {\it The renaissance of General Relativity and Cosmology}, Cambridge Univ. Press 1993
\\
Ericson W., Cronyn W. 1965 ApJ 142, 1156
\\
Esposito S. et al. 2000 {\texttt{astro-ph/0007419}} 
\\
Field G. 1959 ApJ 129, 536
\\
Field G. 1965 ApJ 142, 531
\\
Field G., Somerville W., Dressler K. 1966 Ann. Rev. Astron. Ap. 4, 207
\\
Flower D., Le Bourlot J., Pineau des For\^ets G., Roueff E. 2000 MNRAS 314, 753
\\
Flower D. 2000 MNRAS 318, 875
\\
Flower D., Pineau des For\^ets G. 2000 MNRAS 3116, 901
\\
Galli D., Palla F. 1998 A\&A 335, 403
\\
Gamow G. 1948 Phys. Rev. 74, 505
\\
Gianturco F., Gori-Giorgi P. 1996 Phys. Rev. A 54, 1
\\
Gunn J., Peterson B. 1965 ApJ 142, 1633
\\
Haddock F., Sciama D.W. 1965 Phys. Rev. Lett. 14, 1007
\\
Hannestad S., Raffelt G. 1999 Phys. Rev. D59, 3001
\\
Hannestad S., Hansen S., Villante F. 2000 {\texttt{astro-ph/0012009}} 
\\
Hirasawa T., Aizu K., Taketani M. 1969 Prog. Theor. Phys. 41, 835
\\
Hirasawa T. 1969 Prog. Theor. Phys. 42, 523
\\
Hoyle F. 1953 ApJ 118, 513
\\
Hoyle F. 1958 {\it proceedings of the 11th Solvay congress}
\\
Hu W. et al. 2000 {\texttt{astro-ph/0006436}} 
\\
Hunter C. 1962 ApJ 136, 594
\\ 
Hutchins J. 1976 ApJ 205, 103
\\
Jaffe A. et al. 2000 {\texttt{astro-ph/0007333}}
\\
Jeans J. 1902 Phil. Trans. Roy. Soc. 199A, 49
\\
Jones B., Wyse R. 1985 A\&A  149, 144
\\
Kato S., Nariai H., Tomita K. 1967 Pub. Astron. Soc. Japan 19, 130
\\
Kinney W., Melchiorri A., Riotto A. 2000 {\texttt{astro-ph/0007375}} 
\\
Kompaneets A. 1957, Soviet Phys. JETP 4, 730 
\\
Kondo M. 1970 Pub. Astr. Soc. Japan 22, 23
\\
Kurki-Suonio H. 2000 {\texttt{astro-ph/0002071}} 
\\
Kurki-Suonio H., Sihvola E. 2000 {\texttt{astro-ph/0011544}} 
\\
Lahav O. 1986 MNRAS 220, 259
\\
Latter W., Black J. 1991 ApJ 372, 161
\\
Le Bourlot J., Pineau des Forets G., Flower D. 1999 MNRAS 305, 892
\\
Lepp S., Shull M. 1983 ApJ 270, 578
\\
Lepp S., Shull M. 1984 ApJ 280, 465
\\
Lesgourgues J., Peloso M. 2000 {\texttt{astro-ph/0004412}} 
\\
Lifshitz E. 1946 J. Phys. USSR 10, 116
\\
Lifshitz E., Khalatnikov I. 1963 Advances in Phys. 12, 185
\\
Lisi E., Sarkar S., Villante F. 1999 Phys. Rev. D59, 3520
\\
Mac Dowell M. 1961, Observatory 81, 240
\\
Mac Low M., Shull M. 1986 ApJ 302, 585
\\
Maoli R. et al. 1994 ApJ 425, 372
\\
Maoli R. et al. 1996 ApJ 457, 1
\\
Matsuda T., Sato H., Takeda H. 1969 Prog. Theor. Theor. 41, 219
\\
Melchiorri B., Melchiorri F. 1994 Riv. Nuovo Cim. 17, 1
\\
Melchiorri A., Griffiths L. 2000 {\texttt{astro-ph/0011147}} 
\\
Melott 1984 Soviet Astron. 28, 478.
\\
Mohapatra R., Sciama D.W. 1998 astro-ph/9811446
\\
Novikov I., Zel'dovich Ya. 1967 Ann. Rev. Astr. Ap. 5, 627
\\
Olive K. et al. 1981 ApJ 246, 557
\\
Olive K. et al. 2000 Phys. Rep. 333, 389
\\
Okumurai K. 2000 ApJ 534, 809
\\
Palla F., Salpeter E., Stahler S. 1983 ApJ 271, 632
\\
Parker E. 1953 ApJ 117, 431
\\
Peebles P.J.E. 1966 Phys. Rev. Lett. 16, 410
\\
Peebles P.J.E., Dicke R. 1968 ApJ 154, 891
\\
Puy D. et al. 1993 A\&A 267, 337
\\
Puy D., Signore M. 1996 A\&A 305, 371
\\
Puy D., Signore M. 1997 New Ast. 2, 299
\\
Puy D., Signore M. 1998a New Ast. 3, 27
\\
Puy D., Signore M. 1998b New Ast. 3, 247
\\
Puy D. 2000a A\&A in preparation, see {\texttt{astro-ph/0011496}} 
\\
Puy D. 2000b, in Proceedings {\it 6$^{th}$ Trieste Conference}, 
{\texttt{astro-ph/0011435}} 
\\
Puy D., Signore M. 2001, in {\it Primordial cosmochemistry}, Nova Eds
\\
Rees M., Sciama D.W. 1966 ApJ 145, 6
\\
Rees M., Sciama D.W. 1968 Nature 517, 611
\\
Roberge W., Dalgarno A. 1982 ApJ 255, 489
\\
Rosenzweig P., Parravano A., Ibanez M., Izotov Y. 1994 ApJ 432, 485
\\
Roueff E., Zeippen C. 1999 A\&A 343, 1005
\\
Saito M. 1969 Pub. Astr. Soc. Japan 21, 230
\\
Salati P., Wallet J.C. 1984 Phys. Lett. B 144, 61
\\
Sarkar S. 1996, Rep Prog. Phys. 59, 1493
\\
Sarkar S. 1999, astro-ph 9903183
\\
Saslaw W., Zipoy D. 1967 Nature 216, 967
\\
Schneider D., Schmidt M., Gunn J. 1991 ApJ 102, 837
\\
Schramm D. 1998 Proc. Nat. Acad. Sci USA 95, 42
\\
Schramm D., Turner M. 1998 Rev. Mod. Phys. 70, 303
\\
Sciama D.W. 1964a Quart. Journ. R. Astr. Soc. 5, 196
\\
Sciama D.W. 1964b Nature 201, 767
\\
Sciama D.W. 1982a MNRAS 198, 1
\\
Sciama D.W. 1982b MNRAS 200, 13
\\
Sciama D.W. 1982c Phys. Lett. B 114, 19
\\
Sciama D.W. 1984 in {\it Big Bang: G. Lemaitre in a modern world}, ed Berger A., Dordrecht-Reidel Eds
\\
Sciama D.W. 1988 MNRAS 230, 13p
\\
Sciama D.W. 1990a ApJ 364, 549
\\
Sciama D.W. 1990b Comments Astrophys. 15, 71
\\
Sciama D.W. 1993a in {\it Modern cosmology and the dark matter problem}, 
Cambridge University Press
\\
Sciama D.W. 1993b ApJ 409, L25
\\
Sciama D.W. 1994 Phil. Trans. R. Soc. Lond. A 346, 137
\\
Sciama D.W. 1998 astro-ph/9811172
\\
Sciama D.W. 1999 in: {\it 3K Cosmology}, Eds Maiani et al., 
AIP Conf Proc. 476, 74
\\
Sciama D.W., Rees M. 1966 Nature 212, 1001
\\
Sciama D.W., Melott 1982 MNRAS Phys. Rev. D 25, 2214
\\
Scott D., Silk J., White S. 1995 Science 268, 829
\\
Seager S., Sasselov D., Scott D. 1999 ApJ 523, L1
\\
Seager S., Sasselov D., Scott D. 2000 ApJ 128, 407
\\
Sethi S. 1997 ApJ 474, 13
\\
Shchekinov Y., Ent\'el M. 1983 SvA 27, 622
\\
Shklovskii I 1967, in Proceedings {\it 4$^{th}$ Texas Conf. on Relativistic 
Ap.} 5, 627
\\
Signore M. et al. 1994 ApJ supp 92, 535
\\
Signore M. et al. 1997 Astr. Lett. Com. 35, 349
\\
Signore M., Puy D. 1999 New Ast. Rev. 43, 18
\\
Silk J. 1977 ApJ 211, 638
\\
Sofue Y. 1969 Pub. Astr. Soc. Japan 21, 211
\\
Songaila A., Cowie L., Hogan C., Rugers M. 1994 Nature 368, 599
\\ 
Stancil P., Lepp S., Dalgarno A. 1996 ApJ 472, 102
\\
Stancil P., Lepp S., Dalgarno A. 1998 ApJ 509, 1
\\
Steigman G. et al. 1977 Phys. Lett. B66, 202
\\
Sunyaev R., Zel'dovich Y. 1972 Comments Astroph. Space Phys. 4, 173
\\
Takayanagi K., Nishimura S. 1960 Pub. Astr. Soc. Jap. 12, 77
\\
Takeda H, Sato H. 1968 in {\it Proceedings of symposium on cosmology} 
held at Kyoto February 1968 (in japanese), p 135
\\
Takeda H., Sato H., Matsuda T. 1969 Prog. Theor. Theor. Phys. 41 840
\\
Tegmark M., Silk J., Rees M., Blanchard A., Abel T., Palla F. 1997 ApJ 
474, 1
\\
Tegmark M., Zaldarriaga M. 2000 {\texttt{astro-ph/0004393}} 
\\
Tytler D. et al. 2000 {\texttt{astro-ph/0001318}} 
\\ 
Uehara H., Inutsuka S. 2000 ApJ 531, L91
\\
Wagoner R., Fowler W., Hoyle F. 1967 ApJ 148, 3
\\
Wagoner R. 1969 ApJ supp. 18, 247
\\
Wagoner R. 1973 ApJ 179, 343
\\
Weymann R. 1965 Phys. Fluids 8, 2112
\\
Weymann R. 1966 ApJ 145, 560
\\
White M. et al. 2000 {\texttt{astro-ph/0008167}} 
\\
Yang J. et al. 1979 ApJ 227, 697
\\
Yang J. et al. 1984 ApJ 281, 493
\\
Yoneyama T. 1972 Pub. Astr. Soc. Japan 24, 87
\\
Zel'dovich Y., Kurt V., Sunyaev R. 1968, Zh. Eksp. Teor. Fiz. 55, 287
\\
Zwicky F. 1959 PASP 71, 468
\\
Zygelman B., Stancil P., Dalgarno A. 1998 ApJ 508, 151
}
\end{document}